\documentclass[journal]{ieeecolor}
\usepackage{cite}
\usepackage{amsmath,amssymb,amsfonts}
\usepackage{algorithm}
\usepackage{algpseudocode}
\usepackage{graphicx}
\usepackage{textcomp}
\usepackage{multirow}
\usepackage{fancyhdr}
\usepackage{booktabs}
\usepackage{threeparttable}
\usepackage{subscript}
\usepackage{makecell}
\usepackage{soul,color}
\usepackage{stfloats}
\usepackage{fancyhdr}
\makeatletter
\let\NAT@parse\undefined
\makeatother

\usepackage[colorlinks=true, linkcolor=blue, anchorcolor=cyan, urlcolor=blue,citecolor=blue]{hyperref}

\IEEEaftertitletext{\vspace{-15pt}}
            
\def\BibTeX{{\rm B\kern-.05em{\sc i\kern-.025em b}\kern-.08em
    T\kern-.1667em\lower.7ex\hbox{E}\kern-.125emX}}
    
\begin{document}
\title{Highly Undersampled MRI Reconstruction via a Single Posterior Sampling of Diffusion Models}
\author{Jin Liu, Qing Lin, Zhuang Xiong, Shanshan Shan, Chunyi Liu, Min Li, Feng Liu, G. Bruce Pike,\IEEEmembership{Senior 
Member, IEEE}, Hongfu Sun, Yang Gao
\thanks{YG acknowledges the supported from the National Natural Science Foundation of China under Grant No. 62301616, 62301352, and the Natural Science Foundation of Hunan (Grant No. 2024JJ6530). GBP acknowledges support from the Canadian Institutes for Health Research (CIHR-FDN 143290). HS thanks the support from the Australia Research Council (DE20101297 and DP230101628), and National Health and Medical Research Council of Australia (Grant No. 2030157).}
\thanks{J. Liu, Qi. Lin, M. Li, and Y. Gao are with school of computer science and engineering, Central South University, China.
Z. Xiong is with School of Health Sciences, Faculty of Medicine and Health, University of Sydney, Australia. 
S. Shan is with State Key Laboratory of Radiation, Medicine and Protection, Soochow University, Suzhou, China. 
C. Liu is with Medical School, Nanjing University, China. 
F. Liu is with School of Electrical Engineering and Computer Science, University of Queensland, Brisbane, Australia.
G.B. Pike is with Departments of Radiology and Clinical Neurosciences, Hotchkiss Brain Institute, University of Calgary, Calgary, Canada. 
H. Sun is with School of Engineering, University of Newcastle, Newcastle, Australia. 
Corresponding Author: Yang Gao (\href{mailto:yang.gao@csu.edu.cn}{yang.gao@csu.edu.cn}).
}
}

\maketitle
\begin{abstract}
Incoherent k-space undersampling and deep learning-based reconstruction methods have shown great success in accelerating MRI. However, the performance of most previous methods will degrade dramatically under high acceleration factors, e.g., 8$\times$ or higher. Recently, denoising diffusion models (DM) have demonstrated promising results in solving this issue; however, one major drawback of the DM methods is the long inference time due to a dramatic number of iterative reverse posterior sampling steps. In this work, a Single Step Diffusion Model-based reconstruction framework, namely SSDM-MRI, is proposed for restoring MRI images from highly undersampled k-space. The proposed method achieves one-step reconstruction by first training a conditional DM and then iteratively distilling this model four times using an iterative selective distillation algorithm, which works synergistically with a shortcut reverse sampling strategy for model inference. Comprehensive experiments were carried out on both publicly available fastMRI brain and knee images, as well as an in-house multi-echo GRE (QSM) subject. Overall, the results showed that SSDM-MRI outperformed other methods in terms of numerical metrics (e.g., PSNR and SSIM), error maps, image fine details, and latent susceptibility information hidden in MRI phase images. In addition, the reconstruction time for a  320$\times$320 brain slice of SSDM-MRI is only 0.45 second, which is only comparable to that of a simple U-net, making it a highly effective solution for MRI reconstruction tasks.
\end{abstract}


\fancyfoot[C]{}

\section{Introduction}
\label{sec:introduction}
\IEEEPARstart{M}{agnetic} Resonance Imaging (MRI) is a non-invasive and radiation-free imaging modality that has been widely used in clinical diagnosis and research \cite{frisoni2010clinical,rovira2015magnims} thanks to its excellent tissue contrasts and high image resolutions. However, MRI acquisitions are relatively slow since large areas of k-space need to be traversed, which not only leads to patient discomfort, but also increases the risk of motion artifacts.

Incoherent undersampling combined with compressed sensing (CS) reconstruction algorithms has been substantially studied to accelerate MRI scans, by solving an ill-posed inverse problem. Many traditional iterative reconstruction algorithms have been developed since the pioneering work of Lustig \cite{lustig2007sparse,lustig2008compressed}, such as image sparsity-based CS methods \cite{lustig2007sparse,lustig2008compressed,feng2017compressed,hong2011compressed,doneva2010compressed,yang2014compressed,yang2015aliasing, ong2018general} or low-rankness models \cite{haldar2016p,jin2016general,lingala2011accelerated,zhang2015accelerating}. However, the performance of these algorithms will substantially decrease at high acceleration factors, leading to image blurring or residual artifacts. In addition, they are usually computationally expensive and require manual hyperparameter tuning \cite{ravishankar2019image}. 

Deep learning (DL) algorithms have been increasingly popular as an alternative to conventional iterative methods for MRI acceleration. Many end-to-end supervised deep neural networks \cite{wang2016accelerating,hyun2018deep,shan2023distortion,guo2023reconformer,gao2021accelerating,schlemper2017deep,korkmaz2023self}, such as U-net \cite{ronneberger2015u} based methods, have been proposed for MRI reconstruction recently, demonstrating better artifact removal and improved fine details, compared with traditional iterative methods. However, the generalization capability of these methods is limited, and their performance will degrade when applied to data out of the training distribution. Whereas, unrolling-based or model-based deep neural network \cite{yang2016deep,zhang2018ista,aggarwal2018modl,hammernik2018learning} is another framework incorporating physical models in deep neural network designs, demonstrating better results as well as generalization capability. However, these two frameworks will both decrease their performance when applied to highly undersampled k-space data (e.g., 8$\times$ or higher). 

Recently, diffusion models (DM)\cite{ho2020denoising,song2020score,dhariwal2021diffusion,nichol2021improved, song2020denoising} or noise-conditioned score networks are increasingly popular for MRI reconstruction at high acceleration factors \cite{chung2022score, chu2025highly,jiang2024fast,askari2025training,cao2024high,cui2024spirit,chen2025joint,chung2022come,bian2024diffusion,liu2025score,chung2022diffusion}. This new framework holds great potential for achieving ultra-high k-space acceleration factors, as long as a desirable generative model is successfully trained, which allows high-quality MRI synthesis from only random noise. For example, Score-MRI \cite{chung2022score} successfully learned the time-dependent score functions for high-quality MRI images generation, achieving promising reconstruction quality by iteratively refining undersampled data through stochastic denoising and data-consistency enforcement (at least hundreds of reverse sampling steps required). DiffINR \cite{chu2025highly} proposed a novel posterior sampler for diffusion models using Implicit Neural Representation (INR), which demonstrated better data consistency and more accurate and stable reconstructions at high acceleration rates. It also reduced the necessary number of reverse sampling steps from at least 1000 steps using Score-MRI to only 250 steps using DiffINR. A Predictor-Projector-Noiser (PPN) method \cite{jiang2024fast} was also developed based on Denoising Diffusion Implicit Models (DDIM) \cite{song2020denoising} for controllable and efficient diffusion sampling by repetitively enforcing data fidelity during each generative step. A zero-shot Bi-level Guided Diffusion Model (BGDM) \cite{askari2025training} was developed to solve the inverse problems in medical imaging by firstly adopting an inner-level conditional posterior to obtain an initial prediction and then reinforcing the measurement consistency in the outer-level proximal optimization. High-Frequency Space Diffusion Model (HFS-SDE) \cite{cao2024high} was another framework that was able to accelerate the convergence of the reverse process by confining the diffusion process to high-frequency k-space regions while keeping low-frequency components deterministic, showing improved structural fidelity. SPIRiT-Diffusion \cite{cui2024spirit} chose to incorporate a SPIRiT-based self-consistency prior into the diffusion process, enabling physics-guided reconstruction that enhances robustness under high acceleration factors. More recently, a Come-Closer-Diffuse-Faster (CCDF) \cite{chung2022come} framework was proposed to accelerate the reverse diffusion process of conditional DM models by starting from a forward diffusion point from the initial estimates instead of from pure noise, and it significantly reduces the number of reverse steps to only 20 steps compared to 1000 of Score-MRI. Despite these improvements, a dramatic number of iterative reverse posterior sampling steps are still necessary for these methods to generate reliable and high-quality MRI images, which will accumulate reconstruction times (increasing with more numbers of reverse sampling steps), making the DM-based MRI reconstruction framework much slower than previous DL methods, e.g., end-to-end network-based methods \cite{hyun2018deep,shan2023distortion,guo2023reconformer}. 

In this work, a Single Step Diffusion Model-based method, namely SSDM-MRI, is designed to reconstruct MRI images from highly subsampled k-space in just one posterior sampling step, thus enabling much faster computation than conventional DM-based methods. The proposed method was developed by progressively distilling a pre-trained conditional DM four times using our proposed iterative selective distillation algorithm, which was performed collaboratively with a shortcut reverse sampling strategy to enhance the stability and controllability. Comprehensive experiments were conducted on both the public fastMRI brain and knee datasets \cite{zbontar2018fastmriFa} and an in-house complex-valued brain dataset to compare the proposed SSDM-MRI with multiple established MRI reconstruction methods, including both iterative algorithms and deep learning frameworks. The performances of SSDM-MRI at different acceleration rates were also investigated. In addition, we also evaluated SSDM-MRI's performance on Quantitative Susceptibility Mapping (QSM) acceleration task using the in-house multi-echo brain dataset, and R2* images were also calculated to verify the performance of the proposed SSDM-MRI in multi-echo magnitude imaging contexts. 




\section{Theory and Related Works}                                                                 
\subsection{Problem Formulation}
MRI reconstruction from subsampled k-space data can be formulated as: 
\begin{equation}
\boldsymbol{y}=\boldsymbol{A}\boldsymbol{x}+ \boldsymbol{\epsilon},
\label{eq1}
\end{equation}
where $\boldsymbol{y}\in\mathbb C^{m}$ represents the undersampled k-space measurements and $\boldsymbol{x} \in\mathbb C^{n}$ is the underlying image, and $m \ll n$; $\boldsymbol{\epsilon} \in\mathbb C^{m} $ is the measurement noise, and $\boldsymbol{A} \in \mathbb C^{m \times n} =\boldsymbol{M \times F} $ represents the system forward model, with $\boldsymbol{M} \in \mathbb C^{m \times n}$ being the undersampling mask and $\boldsymbol{F}$ as the Fourier transform. Equation (\ref{eq1}) is known as the general forward model of MRI reconstruction, and it can be used as a data generator to simulate subsampled k-space data from the fully-sampled MRI images, which is popular in many supervised learning frameworks \cite{wang2016accelerating,guo2023reconformer,korkmaz2023self}. In this work, we also adopted Equation (\ref{eq1}) to generate training datasets for network training. 

\begin{figure*}[htbp]  
\centering
\includegraphics[width=0.95\linewidth,scale=0.9]{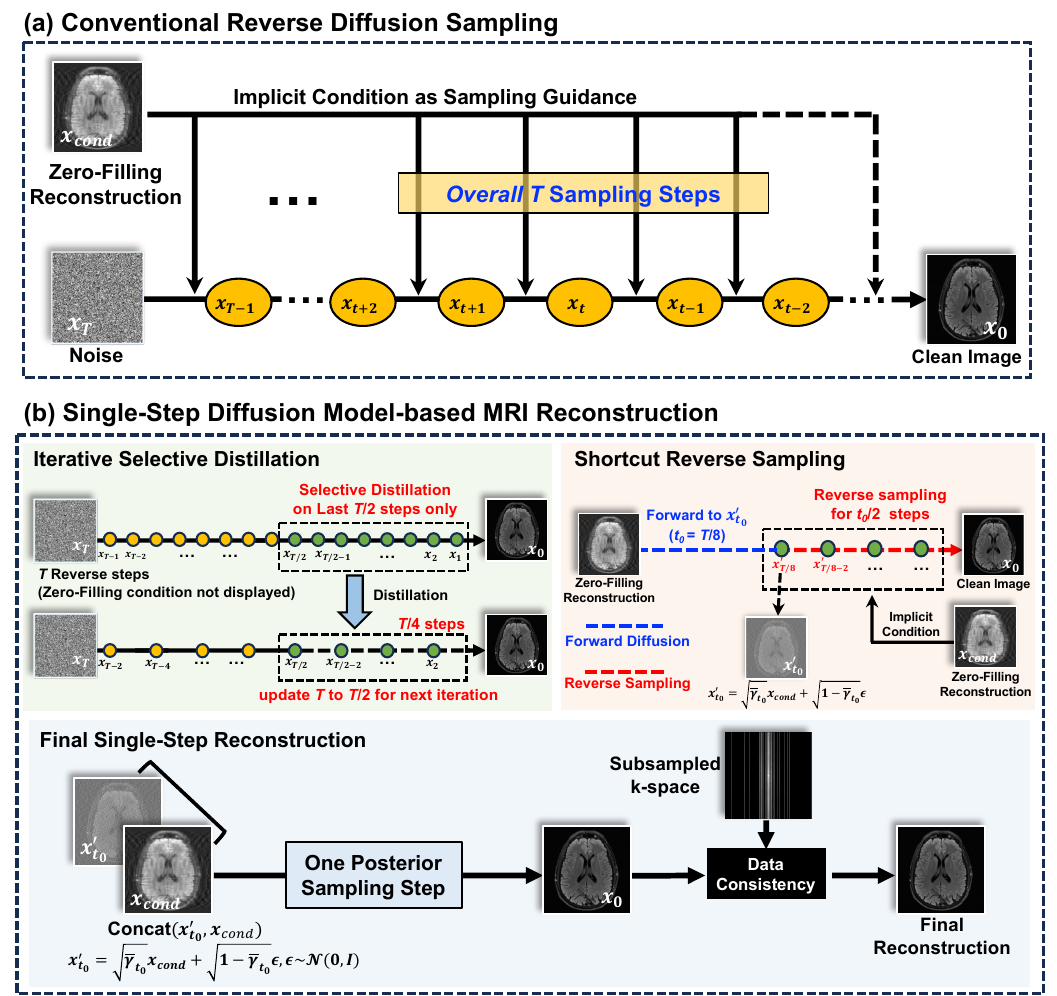}
\caption{Overall framework of the proposed Single-Step Diffusion Model-based MRI reconstruction (SSDM-MRI) method, which is developed by iteratively distilling a pre-trained conditional diffusion model (DM) to reduce the necessary number of reverse sampling steps from original \textit{T} steps to just one step. (a) demonstrates the original \textit{T} reverse steps using the pretrained DM, and (b) illustrates the proposed iterative selective distillation, which is only conducted on the second half \textit{T}/2 steps (green circles) during each iteration, and the paired shortcut reverse sampling strategy, which starts from a single forward diffusion from the zero-filling reconstruction, instead of starting from the pure noise. The bottom panel demonstrates the final single-step reconstruction pipeline after a sufficient number (4 in this work) of iterations of distillation. } 
\label{fig1}
\vspace{-1.0em} 
\end{figure*}

\subsection{Conventional Iterative and Deep Learning Algorithms}
 Equation (\ref{eq1}) could be solved using the following optimization framework, which can be calculated iteratively: 
\begin{equation}
\boldsymbol{x^*}=\underset{ \boldsymbol{x}}{\operatorname{argmin}}  
 f(\boldsymbol{Ax} - \boldsymbol{y})+ \lambda \mathfrak{R}(\boldsymbol{x}),
\label{eq2}
\end{equation}
where $\boldsymbol{x}= \mathfrak{m}e^{i \boldsymbol{\theta}}$; $\mathfrak{m}$ is the magnitude images, and $\boldsymbol{\theta}$ is the phase signals; $f(\cdot)$ is a loss function enforcing data fidelity, e.g., the commonly used \textit{l}2 norm; $\mathfrak{R}(\cdot)$ is a manually-designed regularization term, which can be a sparsifying transformation (e.g., Total Variation and Wavelet Transform), and $\lambda$ is a weighting hyperparameter. In most previous studies \cite{lustig2007sparse,feng2017compressed,hong2011compressed,schlemper2017deep,hammernik2018learning}, the phase signal is simply estimated from the low-frequency part of k-space to simplify this problem; however, a complex phase regularizer is also necessary for phase-based advanced applications \cite{ong2018general,gao2021accelerating}, e.g., Quantitative Susceptibility Mapping (QSM). 

Deep neural networks, due to their excellent performances in complex feature representation capability compared with manually-designed regularizers (i.e., $ \mathfrak{R}(\cdot) $ in (\ref{eq2})), have demonstrated improved reconstruction performances. End-to-end neural networks \cite{hyun2018deep,shan2023distortion,guo2023reconformer} and deep unrolling networks \cite{yang2016deep,zhang2018ista} are two very commonly used frameworks in CS-MRI. The former can be briefly described as follows: 
\begin{equation}
\boldsymbol{\mathfrak{N}_{\theta^*}}=\underset{\theta}{\operatorname{argmin}} \sum_{\mathrm{n}=1}^{\mathrm{N}} f\left( \boldsymbol{x_{\mathrm{n}}}-\boldsymbol{\mathfrak{N}_{\theta}}\left(\boldsymbol{A}^H\boldsymbol{y_{\mathrm{n}}}\right)\right)+\mathfrak{S}(\theta),
\label{eq3}
\end{equation}
where $\boldsymbol{y_n}$ and $\boldsymbol{x_n}$ denote the $n^{th}$  training sample pair of sub-sampled k-space data and the corresponding fully-sampled image in the training dataset, respectively, and $\theta$ is the learnable parameters in the deep network $\boldsymbol{\mathfrak{N}}$, and $f(\cdot)$ is the loss function adopted for network training; $\mathfrak{S}(\theta)$ represents the techniques used to address the potential overfitting problems in deep and large models. In this scheme, end-to-end networks tried to directly learn the non-linear mapping between training inputs, which is usually the zero-filling reconstructions (i.e., $\boldsymbol{A}^H \boldsymbol{y_n}$), and training labels ($\boldsymbol{x_n}$). Whereas deep unrolling networks, on the other hand, are usually constructed by iteratively stacking physical-model modules and neural networks. The network part usually tries to learn the mapping between latent variables in conventional iterative algorithms (i.e., learning the proximal gradient descent process in ISTA-Net \cite{zhang2018ista}). 

\subsection{Diffusion Models for MRI Reconstruction}
Diffusion models (DM)\cite{ho2020denoising,dhariwal2021diffusion,nichol2021improved} or the noise-conditioned score networks \cite{song2020score,song2020denoising} are new types of generative models that are capable of synthesizing high-quality images via gradually denoising Gaussian noise. It involves a forward noise adding process to convert the original image into pure Gaussian noise, and an iterative backward sampling process to recover the original images in $T$ consecutive steps. DM is naturally suitable for solving inverse problems and has shown great potential in MRI reconstructions \cite{chu2025highly,bian2024diffusion,chung2022score,cao2024high,liu2025score,chung2022diffusion,cui2024spirit,chen2025joint}. 

Suppose that the desired MRI images follow a specific unknown distribution (i.e., $\boldsymbol{x}\sim p(\boldsymbol{x})$), then the reconstruction of MRI images ($\boldsymbol{x}$) from subsampled k-space ($\boldsymbol{y}$) can be formulated as a posterior sampling process $p(\boldsymbol{x}|\boldsymbol{y})$, and the key to obtain reliable samples from this conditional distribution in the DM backward process is to estimate the corresponding score function $\boldsymbol{\nabla_{\boldsymbol{x_t}}\log(p(\boldsymbol{x_t}|\boldsymbol{y}))}$ \cite{song2020score,song2021solving} (i.e., the noise estimation in DM). One can train a deep neural network to estimate this conditional score ($\boldsymbol{\mathfrak{N}_{\theta}(x_t, y) \approx \nabla_{\boldsymbol{x_t}}\log(p(\boldsymbol{x_t}|\boldsymbol{y})}$) in a supervised or unsupervised manner. 



\subsection{Progressive Diffusion Model Distillation} 
One major drawback of DM-based methods is the slow (reverse) sampling process, which usually takes a large number of posterior sampling steps. Distillation strategies were commonly employed to accelerate the posterior sampling process in many previous works \cite{song2020denoising,song2023consistency}, among which progressive distillation \cite{salimans2022progressive} has been one of the most effective techniques. It iteratively reduces the number of necessary sampling steps by distilling a teacher diffusion model into a faster student model, which then becomes a new teacher for the next iterative distillation. Every distillation reduces the number of necessary sampling steps from $T$ steps to $T/2$ steps. 

\subsection{CCDF: Accelerating the Sampling of Conditional Diffusion Models}
Come-Closer-Diffuse-Faster (CCDF) \cite{chung2022come} is an efficient and effective framework for accelerating the reverse sampling of pretrained conditional diffusion models for inverse problems. It is shown that there exists a shortcut reverse sampling path for a pretrained conditional DM, so that the inference can start from a closer point and achieve improved final results in much fewer sampling steps. The key design is to start the inference of DMs with a forward diffusion of the initial estimates of the desired images, instead of from pure Gaussian noise. As report in the original work\cite{chung2022come}, 20 reverse sampling steps are enough for CCDF to obtain more accurate reconstructions than Score-MRI \cite{chung2022score} of 1000 reverse steps. 

\section{Method}         
\subsection{Overall Framework of SSDM-MRI}
The proposed SSDM-MRI was constructed by two steps: (\textbf{1}), training a conditional denoising diffusion probability model (DDPM) \cite{ho2020denoising} in a supervised manner, which required $T$ reverse sampling steps to generate clean images ($\boldsymbol{x_0}$) from pure Gaussian noise ($\boldsymbol{x_T}$) and the zero-filling image ($\boldsymbol{x_{cond}}=\boldsymbol{A}^H \boldsymbol{y_n})$, i.e., the generation condition), as shown in Fig. \ref{fig1}(a), and (\textbf{2}), distilling the pretrained DM using the proposed iterative selective distillation algorithm, which was performed in pairs with the shortcut reverse sampling strategy. The proposed selective distillation algorithm and the shortcut reverse sampling strategy are detailed in \textbf{Subsection \ref{subsection-ISD}}. 

As shown in the bottom panel of Fig. \ref{fig1}(b), the proposed SSDM-MRI takes the zero-filling image ($\boldsymbol{x_{cond}}$) and a latent image ($\boldsymbol{x_{t_0}'}$ detailed in \textbf{Subsection \ref{subsection-ISD}}) forward diffused from the zero-filling image as inputs (concatenated along the channel dimension), and generates desired MRI images as outputs. To achieve more robust and stable image reconstruction, a projection-based data consistency \cite{schlemper2017deep,gao2021accelerating} was also performed to further refine the DM-synthesized images to impose the k-space data fidelity, as described in (\ref{eqdc}) below:

\begin{equation}
\boldsymbol{x_{dc}(k) = (1 - M) \cdot x_0(k) + y(k)}, 
\label{eqdc}
\end{equation}
where $\boldsymbol{x_{0}(k)}$ is the original diffusion model reconstructions in k-space, $\boldsymbol{M}$ is the undersampling mask, and $\boldsymbol{y(k)}$ is the subsampled k-space data. The final reconstruction of SSDM-MRI is the inverse Fourier Transform of k-space signal $\boldsymbol{x_{dc}(k)}$. 




\begin{algorithm}[t]
\caption{Conditional Diffusion Model Pre-Training}
\begin{algorithmic}[1]
\Require $x_0 \sim$ fully sampled images
\Statex \hspace{2.1em} $x_{cond} \sim \text{Zero Filling Reconstruction} (A^Hy)$
\Statex \hspace{2.1em} Network $\mathfrak{N}_{\theta}$ to be optimized 
\While{not converged}
    \State $t \sim DiscreteU({1, T})$
    \State $\varepsilon \sim \mathcal{N}(0, I)$
    \State $\alpha_t = \sqrt{\bar{\gamma_t}},\ \sigma_t = \sqrt{1 - \bar{\gamma_t}}$
    \State $v=\alpha_t \varepsilon- \sigma_t x_0$
    \State Gradient descent on: 
    \Statex \hspace{4.5em} $\nabla_\theta \left\| \mathfrak{N}_{\theta}( x_{cond},\ \alpha_t x_0 + \sigma_t \varepsilon,\ \bar{\gamma_t} ) - v \right\|$
\EndWhile
\end{algorithmic}
\end{algorithm}
\setlength{\textfloatsep}{1pt}

\subsection{Conditional Diffusion Model Pretraining}
Similar to most previous works \cite{ho2020denoising,song2020score}, the forward process is a stable Markov chain of diffusion steps, gradually adding Gaussian noise into the fully-sampled MRI images ($\boldsymbol{x_0}$), which is controlled by a parameter scheduler $\left\{ \boldsymbol{\gamma_{t} \in (0,1)},t = {1,2,3..,T} \right\}$. In this work, $T$ was set as 128 and a cosine noise scheduler ($\gamma_t = cos^2(\pi/2 *t/T)$) was adopted during DM pretraining, and the forward process can be expressed as follows: 
\begin{equation}
\begin{aligned}
\boldsymbol{x_{t}}=\sqrt{\bar{\gamma_{t}}} \boldsymbol{x_{0}}+ \sqrt{1 - \bar{\gamma_{t}}} \boldsymbol{\epsilon}, \quad \boldsymbol{\epsilon} \sim \mathcal{N}(0, \mathbf{I}) ,
\end{aligned}
\label{eq4}
\end{equation}
where $x_t$ is the noise contaminated image at time step $t$, $x_0$ is the noise free images, $x_T$ is pure Gaussian noise, and $\bar{\gamma_{t}}=\prod_{i=1}^{t} \gamma_i$. 

The pre-trained conditional DM was obtained by training a self-tailored residual U-net ($\boldsymbol{\mathfrak{N}_{\theta}}$) on a large number of paired datasets ($\left\{ \boldsymbol{(A^Hy)^{n}}, \boldsymbol{x_0^{n}} \right\}_{n=1,2,3...N}$) simulated based on (\ref{eq1}), as detailed in \textbf{Algorithm 1}. Overall, the U-net takes both $\boldsymbol{x_t}$ and zero-filling reconstructions ($\boldsymbol{A^Hy}$) as its major inputs (concatenated along the channel dimension). Similar to most previous U-net models, this adopted U-net consists of an encoding and a decoding path, each consisting of three modules, and each module contains three convolutional blocks with residual connections from the block input to the output. Three pooling layers and three corresponding unpooling layers were included for multiscale or multi-resolution learning. Self-attention operations were only carried out on the latent features of the lowest resolution due to memory constraints. Overall, the network consists of 52 convolutional layers (kernel size: 3$\times$3), 3 max-pooling layers (kernel size: 2$\times$2), 3 transposed convolutional layers (kernel size: 2$\times$2), 51 batch-normalization layers, 77 rectified linear units (ReLUs), 3 concatenations, 25 residual connections, and 6 self-attention operations. The network also takes the noise scheduler ($\bar{\gamma_{t}}$) as its third input, which was encoded using a simple thee-layer multi-layer perceptron, inspired by previous works \cite{saharia2022palette} focusing on conditional generation. 

In this work, the network is designed to predict a latent variable $\boldsymbol{v =\sqrt{\bar{\gamma_t}}\epsilon - \sqrt{1-\bar{\gamma_t}}x_0}$, which is better suited for distillation \cite{salimans2022progressive}, considering the fact that the standard noise prediction parameterization used in DDPM is not well suited for distillation, particularly as the signal-to-noise ratio (SNR) approaches zero, where small errors in noise prediction can be significantly amplified, leading to unstable training and degraded performance. Using the v-diffusion parameterization, the denoising process turns into $\boldsymbol{\hat{x}_0=\alpha_t x_t-\sigma_t \mathfrak{N}_{\theta}}$, which better preserves stable supervision signals during distillation and enables more efficient sampling. Following this modification, the proposed iterative selective distillation is then carried out, and in each distillation iteration, the student model is optimized to replicate the behavior of two sampling steps of the teacher model in just one step, as detailed in \textbf{Algorithm 2} (line 7-21).

\setlength{\textfloatsep}{6pt}
\begin{algorithm}[t]
\caption{Iterative Selective Distillation}
\begin{algorithmic}[1]
\Require $\mathfrak{N}_{\eta}$: pre-trained teacher model ($T$ reverse steps) 
\Statex \hspace{2.5em}$\mathfrak{N}_{\theta}$: student model
\Statex \hspace{2.5em}$x_0 \sim$ fully sampled images
\Statex \hspace{2.5em}$x_{cond} \sim \text{Zero Filling Reconstructions} (A^Hy)$
\Statex \hspace{2.5em}$DC:$ Data Consistency
\For{$k = {1, 2, ... K}$ }
    \State Initialize $\mathfrak{N}_{\theta}$ from $\mathfrak{N}_{\eta}$
    \State $T'= T / 2$ \Comment{selective distillation on last $T / 2$ steps}
    \While{not converged}
        \State $\varepsilon \sim \mathcal{N}(0, I)$
        \State $t \sim  \mathrm{DiscreteU}(1, T'/2) \times 2$ 
        \State $\alpha_t = \sqrt{\bar{\gamma}_t},\ \sigma_t = \sqrt{1 - \bar{\gamma}_t}$
        \State $t_1 = t-1,\quad t_2 = t-2$  
        \State $x_t = \alpha_t x_0 + \sigma_t \varepsilon$
        \State $x_{0}'=\alpha_t x_t -\sigma_t \mathfrak{N}_{\eta}(x_{cond},x_t,\bar{\gamma}_t)$ 
        \State $x_{t_1} = \alpha_{t_1} x_{0}' + \frac{\sigma_{t_1}}{\sigma_t} \left( x_t - \alpha_{t} x_{0}' \right)$
        \\ \text{\quad \quad \quad \#\# two steps by teacher } 
        \State $x_{0}''=\alpha_{t_1} x_{t_1} -\sigma_{t_1} \mathfrak{N}_{\eta} 
        (x_{cond},x_{t_1},\bar{\gamma}_{t_1})$ 
        \State $\varepsilon_{t_2}=\frac{(x_t-\alpha_{t_2} x_{0}'')}{\sigma_{t_2}}$
        \State $\hat{v} = \alpha_{t_2} \varepsilon_{t_2}-\sigma_{t_2} x_{0}''$
         \State $v=\mathfrak{N}_{\theta}(x_{cond}, x_t, \bar{\gamma}_{t})$ 
          \State $x_{0}'''=\alpha_{t} x_{t} -\sigma_{t}v$   \Comment{one step by student }
        \State $x_{teacher}=\mathrm{DC}(x_{0}''), x_{student}=\mathrm{DC}(x_{0}''')$
        \State $L_{1} = \left\| \hat{v}-v  \right\|_2^2$
        \State $L_{2} = \left\| x_{teacher}-x_{student} \right\|_2^2$
        \State $L_{\theta}=L_{1}+L_{2}$  
    \EndWhile
    \State $\mathfrak{N}_{\theta} \Rightarrow \mathfrak{N}_{\eta}$  \Comment{Make student next teacher}
    \State $T \rightarrow T/2$  \Comment{Sampling steps halved}
\EndFor
\end{algorithmic}
\end{algorithm}
\setlength{\textfloatsep}{1pt}


\subsection{Iterative Selective Distillation and Shortcut Reverse Sampling}
\label{subsection-ISD}
Fig. \ref{fig1}(b) illustrates the main concepts of the proposed iterative selective distillation algorithm (the green block) and a paired shortcut reverse sampling strategy (the orange block), which were conducted synergistically to reduce the reverse sampling steps of the pretrained DM from the original $T$ steps to only one step (blue block). The implementation details of the proposed iterative selective distillation are depicted in \textbf{Algorithm 2}, which is inspired by and modified from the progressive distillation algorithm \cite{salimans2022progressive}. The major improvement is that the current selective distillation strategy only carries out the distillation process on the latter $T/2$ reverse sampling steps (from $x_{T/2}$ to clean image $x_0$, i.e., the green circles in the dashed boxes in Fig. \ref{fig1}(b)), as also indicated in line 3 and 6 of \textbf{Algorithm 2}, while the conventional progressive distillation strategy is performed on the entire reverse sampling path (all $T$ reverse sampling steps). Furthermore, the proposed \textbf{Algorithm 2} also incorporated a DC operation on the latent results of the teacher and student models, i.e., $x_0''$ and $x_0'''$ in line 18, and a second loss was introduced on the DC results in line 20. 

As the first half reverse sampling path (from pure noise $x_{T}$ to $x_{T/2}$) is ignored during the distillation, the sampling method of the proposed SSDM-MRI was updated correspondingly in this work. Inspired by CCDF \cite{chung2022come}, a shortcut reverse sampling strategy was adopted to start the reverse sampling of the pretrained DM from a latent $\boldsymbol{x_{t_0}'}$ point instead of pure noise ($x_{T}$), which was calculated by a forward diffusion on the zero-filling reconstruction (the blue dashed line in Fig. \ref{fig1}(b)) using the following equation: 
\begin{equation}
\begin{aligned}
\boldsymbol{x_{t_0}'}=\sqrt{\bar{\gamma}_{t_0}} \boldsymbol{x_{cond}}+ \sqrt{1 - \bar{\gamma}_{t_0}} \boldsymbol{\epsilon}, \quad \boldsymbol{\epsilon} \sim \mathcal{N}(0, \mathbf{I}),
\end{aligned}
\label{eq-shortcut}
\end{equation}
where $\boldsymbol{x_{cond}}=\boldsymbol{A}^H \boldsymbol{y}$ is the zero-filling reconstructions from the subsample k-space data ($\boldsymbol{y}$); $\bar{\gamma}_{t_0}=\prod_{i=1}^{t_0} \gamma_i$ as defined in \ref{eq4}. $t_0$ is a latent point on the reverse sampling path of the pretrained DM, and it is empirically set as $T/8$ in this work. 

Combining the iterative selective distillation and shortcut reverse sampling, the numbers of the reverse sampling steps for the distilled models after the $k^{th}$ distillation stage is reduced from the original $T$ steps to $T/8 \times (1/2)^{k}$ steps only, which is much more efficient than the conventional progressive distillation, where the resulting model of $k^{th}$ distillation will need $T\times (1/2)^k$ steps for high quality image generation. Therefore, for the pretrained conditional DM in the previous section ($T=128$), only four iterative distillations (i.e., set $K=4$ in \textbf{Algorithm 2}) are enough to obtain a single-step model for the proposed SSDM-MRI. In addition, as the student model's performance would decrease more with an increasing number of distillations, compared with the pretrained model, there would be fewer performance declines using the proposed iterative selective distillation and shortcut reverse sampling framework. Further, the proposed design is intrinsically beneficial to improve the model's performance because the student model during model distillation now only needs to learn half of the dynamics in the reverse diffusion path, rather than the entire reverse path from the pure Gaussian noise to clean images.

\begin{figure*}[t]  
\centering
\includegraphics[width=.95\linewidth,scale=0.8]{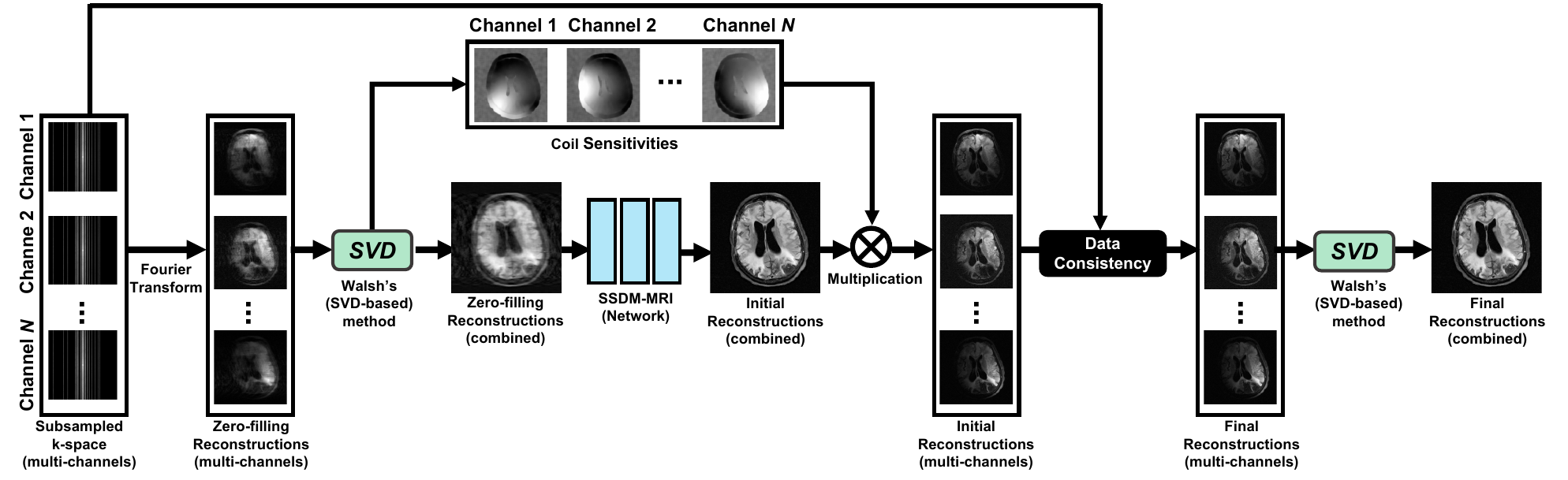}
\caption{The reconstruction pipeline of the proposed SSDM-MRI for multi-coil data. The Walsh’s (SVD-based) method was used to estimate the coil sensitivity maps from the zero-filled images in image space and combine the multi-channel data. }
\label{fig2}
\vspace{-1.0em} 
\end{figure*}

\subsection{SSDM-MRI for Multi-Coil Data and QSM Acceleration}
As shown in Fig. \ref{fig2}, the reconstruction of SSDM-MRI for multi-coil data can be carried out using the following pipeline: (1) zero-filling reconstructions was first performed for each receiver channel, (2) then, the coil sensitivity maps were estimated using Walsh’s (SVD-based) method \cite{walsh2000adaptive} for coil combination, because it can estimate the coil sensitivities in image-domain without explicitly specifying ACS lines, (3) next, the combined images were fed into SSDM-MRI network (without data consistency) for initial reconstruction, (4) then the initial reconstructions were multiplied by the estimated coil sensitivities to convert back into a multi-channel k-space, (5) next, data consistency was conducted for every coil to obtain images for each coil channel, (6) finally, these multi-coil images were re-combined using Walsh’s (SVD-based) method \cite{walsh2000adaptive}. 

Although the proposed SSDM-MRI is designed and trained for 2D MRI image reconstructions, it can be easily transferred to advanced 3D MRI-based applications in a slice-by-slice manner, following the strategy proposed in our previous QSM acceleration work \cite{gao2021accelerating}, where the 3D subsampled k-space volumes are first converted into 2D k-space slices after a 1D Fourier Transform along the fully-sampled (frequency-encoding) direction before being fed into the networks for reconstructions. 3D QSM images can then be calculated from the reconstructed MRI phase images, using best path phase unwrapping \cite{abdul2007fast,abdul2009robust}, RESHARP-based background removal \cite{sun2014background}, and COSMOS dipole inversion \cite{liu2009calculation}. 

In this work, the 2D subsampling patterns involved were only applied to 3D k-space volumes to make sure that the readout (frequency-encoding) direction is always fully sampled to meet the MRI acquisition convention. 



\vspace{-1.0em} 
\subsection{Training Data Preparation and Network Training}
In this work, two different instances of the proposed SSDM-MRI were obtained by using two different types of training datasets, i.e., (1) the public fastMRI brain dataset \cite{zbontar2018fastmriFa}, and (2) an in-house multi-echo gradient echo (mGRE) complex-valued MRI dataset used in our recent work \cite{gao2021accelerating}. In this study, except for the QSM-oriented comparative study, all other evaluations were carried out using the one trained on fastMRI data. 

Details about these two datasets are as follows: 

\textbf{(1) fastMRI Brain Dataset} We extracted a total of 34,063 2D brain images (size: 320$\times$320) from 2,610 volumes randomly selected from the \href{https://fastmri.org}{fastMRI} brain train dataset batch \cite{zbontar2018fastmriFa}. The last three slices of each volume were discarded to avoid noise-only slices. All the images were linearly normalized (divided by the maximum intensity of the image) before being fed into the network for training . Different undersampling masks of various types (including 1D and  2D Gaussian masks, and 2D Poisson masks under various AFs, i.e., 4$\times$, 6$\times$, 8$\times$, 10$\times$,  12$\times$, and 15$\times$) were randomly simulated at the beginning of every training mini-batch to synthesize subsampled k-space data and the corresponding zero-filling reconstructions.

\textbf{(2) mGRE Brain Data} A total of 44,063 complex-valued slices (image size: 256$\times$256) were retrospectively obtained from 47 subjects (Institutional ethics board approvals have been obtained for all MRI-QSM brain data used in this work), which were acquired at 3T (Discovery 750, GE Healthcare, Milwaukee, WI), with the following parameters: 8 unipolar echoes, first TE / $\Delta$TE = 3 ms / 3.3 ms, TR = 29.8 ms, FOV = 256 mm $\times$ 256 mm $\times$ 128 mm, resolution = 1 mm isotropic. The image normalization and subsampled k-space simulation were made the same as the pipeline described for fastMRI brain dataset. 

All network training were conducted on two NVIDIA V100 (32 GB) GPUs, based on the loss functions defined in \textbf{Algorithm 1 and 2}. For DM pretraining, all learnable parameters in the network were initialized as random numbers with a mean of 1 and standard deviation of 0.001, and are optimized using the Adam optimizer with a learning rate of 5 $\times 10^{-5}$ for 200 epochs. The proposed models were implemented using Pytorch 1.13.1. The source codes and the trained models are available at \href{https://github.com/YangGaoUQ/SSDM-MRI}{https://github.com/YangGaoUQ/SSDM-MRI}. 

\section{Experiments}
\subsection{Evaluation Datasets}
The performance and generalization capability of the proposed SSDM-MRI were comprehensively evaluated on the following retrospectively and prospectively undersampled data, including:
\begin{enumerate}
\item[(1)] 300 in distribution single-coil brain slices from fastMRI at acceleration factors (AF) of 8$\times$ and 12$\times$ (1D Gaussian subsampling, ACS = 12 and 6 lines, respectively), and 15$\times$  (2D Poisson). 
\item[(2)] 300 Out-of-distribution (OOD) fastMRI knee dataset at AF of 8$\times$ (1D Gaussian) and 12$\times$ (2D Gaussian). 
\item[(3)] A single-coil pathological slice containing brain lesions (AF = $8\times$), simulated from a subject acquired at 3T using the parameters of the mGRE brain training dataset.
\item[(4)] Retrospective 12-coil brain (320 images at AF = 12$\times$, 1D Gaussian and random mask) and 15-coil knee (320 images at AF = 12$\times$ of 1D Gaussian masks and 15$\times$ of 2D Poisson masks) datasets with ground truths.
\item[(5)] Two 16-coil prospectively undersampled brain images (AF = $8\times$ and $4\times$) randomly selected from fastMRI test dataset.
\item[(6)] A brain subject of five repetitive mGRE scans (size: $256 \times 256 \times128 \times8$) acquired at different orientations, i.e., neutral head position (left 23°, right 17°, flexion 18°, and extension 21° tilt angles from the neutral head position) retrospectively undersampled at AF = 8 using sampling masks adopted in \cite{gao2021accelerating} to evaluate SSDM-MRI's performance on advanced QSM acceleration task. COSMOS-QSM images were reconstructed based on the five repetitive scans of the mGRE brain subject at all five different head orientations, and the magnitude-based R2* images were calculated using the data of neutral head orientation. 

\end{enumerate}

\subsection{Performance Evaluation and Numerical Metrics}

\begin{figure*}[t]  
\centering
\includegraphics[width=.95\linewidth,scale=0.8]{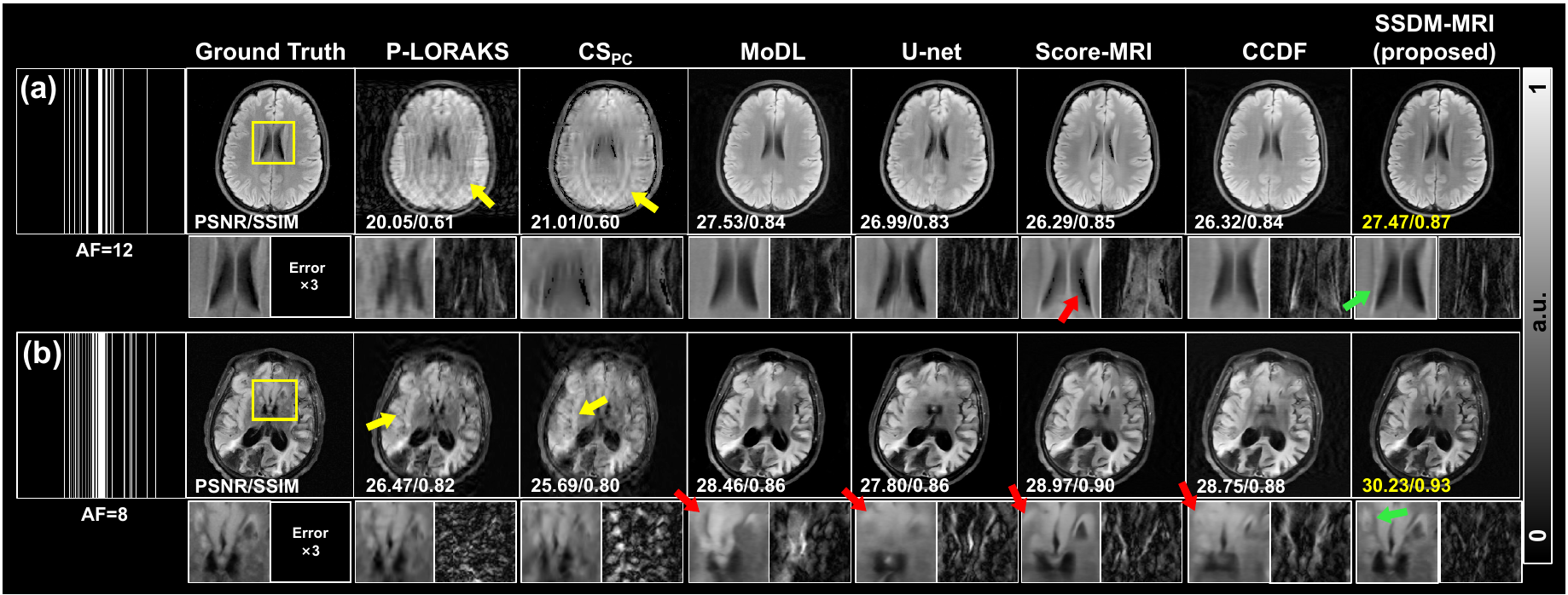}
\caption{Comparison of SSDM-MRI with various reconstruction methods on two single-coil brain slices. (a) and (b) demonstrates the results under acceleration factors of 12 (1D Gaussian, ACS = 6 lines) and 8 (1D Gaussian, ACS = 12 lines), respectively. The yellow arrows point to the residual artifacts in P-LORAKS, CS\textsubscript{PC}, and the red arrows point to the reconstruction errors of fine structures in MoDL, U-net, Score-MRI, and CCDF, while green arrows in the zoomed-in image point to improved fine details in SSDM-MRI.}
\label{fig3}
\vspace{-1.0em} 
\end{figure*}

\begin{figure*}[b]  
\centering
\includegraphics[width=.95\linewidth,scale=0.9]{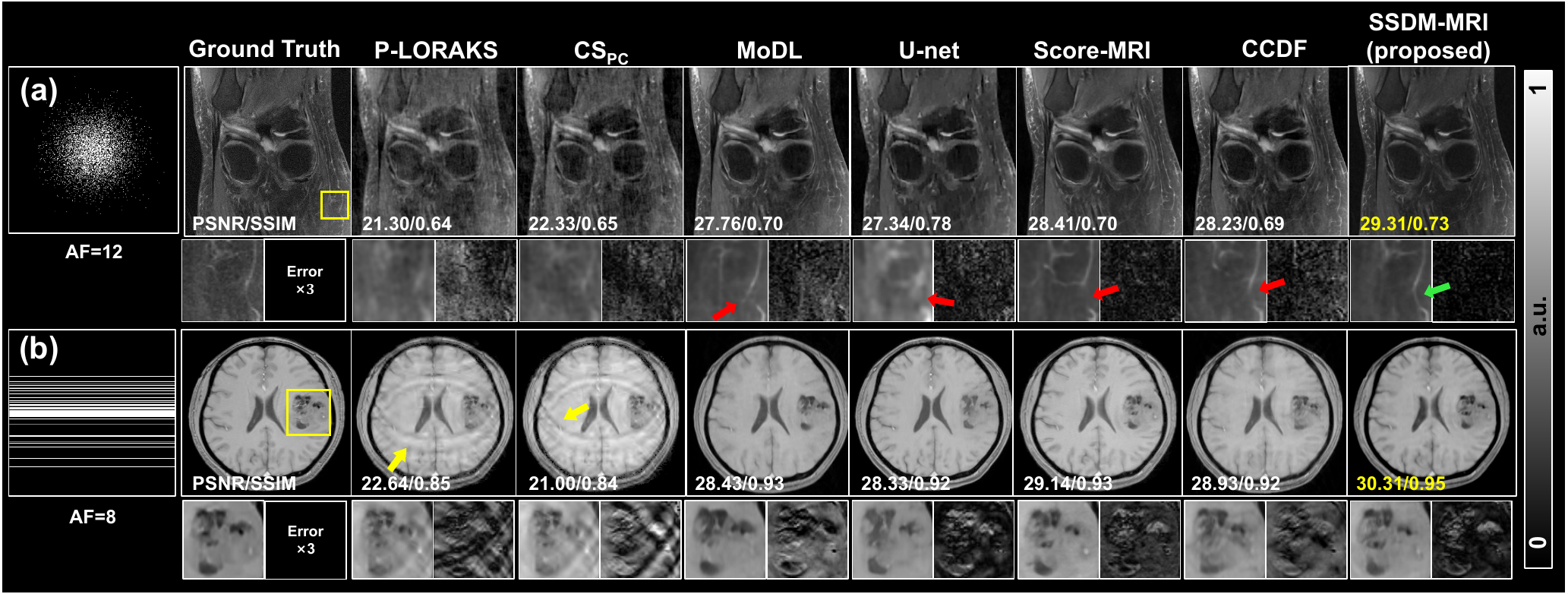}
\caption{Comparison of SSDM-MRI with P-LORAKS, CS\textsubscript{PC}, MoDL, U-net, Score-MRI, and CCDF on two slices from (a) fastMRI knee volume at AF = 12 and (b) a pathological subject with severe brain lesions of AF = 8, respectively. Zoomed-in regions and the corresponding error maps were reported below the reconstruction results. The yellow arrows point to the residual artifacts in P-LORAKS, CS\textsubscript{PC}, and the red arrows point to the errors of MoDL, U-net, Score-MRI, and CCDF. The green arrow points to better image fine details in SSDM-MRI. } 
\label{fig4}
\vspace{-1.0em} 
\end{figure*}




SSDM-MRI was compared with the two traditional algorithms and four deep learning methods: (1) scan-specific P-LORAKS \cite{haldar2016p} (LORAKS type = “S”, rank = 40, and maximum iteration number = 100), (2) Compressed Sensing with Phase Cycling (CS\textsubscript{PC}) \cite{ong2018general} ($\lambda_m$  = 0.003, $\lambda_p$  = 0.005, inner iteration number = 10, and out iteration number = 100), (3) the unrolling-based MoDL\cite{aggarwal2018modl} method, (4) end-to-end U-net based method \cite{ronneberger2015u}, and (5) DM-based Score-based MRI \cite{chung2022score} ($T=1000$ overall sampling steps), and (6) CCDF \cite{chung2022come} (20 reverse steps as set in the original work) on the above described datasets of different acceleration factors and subsampling patterns. The data consistency operation was applied to all the methods involved in this work for fair comparisons. The deep learning methods were re-trained using the same datasets that were 
originally used for training the SSDM-MRI. The reconstruction times of different methods on a image of size $320\times320$ are summarized in Table \ref{table1}, indicating that the reconstruction speed of SSDM-MRI is much faster than that of Score-MRI, since only one reverse step is needed. 

\begin{figure*}[t]  
\centering
\includegraphics[width=0.95\linewidth,scale=0.8]{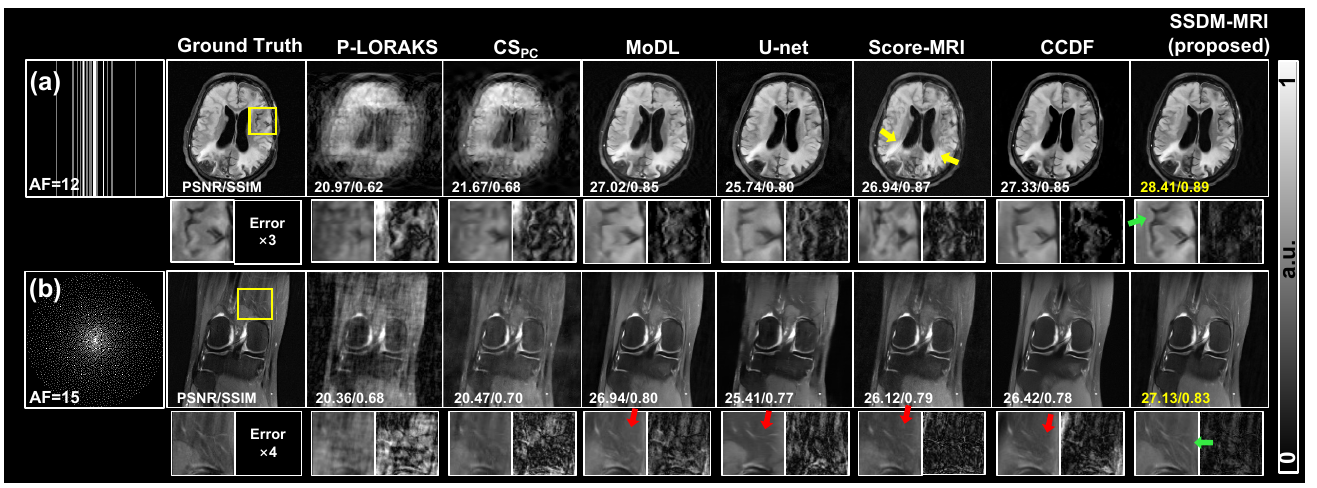}
\caption{Comparison of different MRI accelerating methods on (a) a 12-coil brain slice and (b) a 15-channel knee image from fastMRI at AF = 12 and 15, respectively. The red arrows point to detail loss in MoDL, U-net, Score-MRI, and CCDF in the zoomed-in region, while the green arrows point to better fine details preserved in the proposed SSDM-MRI.}
\label{fig5}
\vspace{-0.7em} 
\end{figure*}

\begin{table}[t]
\begin{threeparttable}
\centering
\caption{Reconstruction times of different methods on a brain image of 320 $\times$  320 size. }
\renewcommand{\arraystretch}{1}
\setlength{\tabcolsep}{1.5pt}
\begin{tabular}{ccccccc}
\toprule
 \textbf{P-LORAKS} & \textbf{CS\textsubscript{PC}} & \textbf{MoDL} & \textbf{U-net} & \textbf{Score-MRI} & \textbf{CCDF} & \textbf{SSDM-MRI} \\
\midrule
2.67 s    & 221.45 s  &1.91 s     &0.36 s  & 315.18 s & 7.45 s & 0.45 s \\
\bottomrule
\end{tabular}

\begin{tablenotes}
\footnotesize{
\item \textbf{P-LORAKS} and \textbf{CS\textsubscript{PC}} were run on a Intel i5-12400 CPU Processor, while the other deep learning method were evaluated on a NVIDIA 3060 GPU. }
\end{tablenotes}
\label{table1}
\end{threeparttable}
\end{table}

Performance evaluation experiments are carried out as follows: (1) First, the proposed SSDM-MRI was compared with other reconstruction methods using the single-coil fastMRI brain dataset to demonstrate its superior performance on the in-distribution data. (2) Subsequently, it was tested on OOD fastMRI knee images and the simulated pathological lesion brain image to test its generalization capability. (3) Furthermore, SSDM-MRI’s performances were further validated on both the retrospective and prospective multi-coil fastMRI data to further demonstrate its generalization capability. (4) In addition, SSDM-MRI’s performance was evaluated on one 8-echo GRE brain subject for QSM acceleration following \cite{gao2021accelerating}, which could compare various methods at the whole brain level instead of a single 2D slice, and evaluate their capability of preserving latent quantitative susceptibility information. This experiment can also compare the performance of SSDM-MRI with other methods from the perspective of MRI phase image reconstruction, which is often neglected in previous works. R2* image-based comparisons were also conducted to explore SSDM-MRI’s applicability in multi-echo magnitude imaging contexts. (5) One more simulated study was carried out to investigate the performance of SSDM-MRI against various acceleration factors (AF = 4$\times$, 6$\times$, 8$\times$, 10$\times$, 12$\times$, respectively). (6) Moreover, the stability of SSDM-MRI was compared with Score-MRI by inspecting the generation of uncertainty maps and a quantitative uncertainty comparison of the proposed SSDM-MRI with the other two DM methods, i.e., Score-MRI and CCDF. (7)  Finally, an ablation study was performed to investigate the effectiveness of the proposed iterative selective distillation algorithm (paired with the proposed shortcut reverse sampling), the number of distillation iterations, and the data consistency (DC) module on the reconstruction result. 


\begin{table*}[b]

    \centering
    \caption{Comparison of different MRI methods on the four retrospective fastMRI datasets in terms of PSNR and SSIM (the best results are highlighted in bold).}

\begin{threeparttable}
    
\resizebox{\textwidth}{!}{%
\begin{tabular}{llcccccccccccccccc}
        \toprule
        \multirow{2}{*}{\normalsize\hspace{1.3em} Dataset} & \multirow{2}{*}{\normalsize\hspace{0.8em} Mask}
        & \multicolumn{1}{c}{P-LORAKS} 
        & \multicolumn{1}{c}{CS\textsubscript{PC}} 
        & \multicolumn{1}{c}{MoDL} 
        & \multicolumn{1}{c}{U-net} 
        & \multicolumn{2}{c}{Score-MRI} 
        & \multicolumn{2}{c}{CCDF} 
        & \multicolumn{1}{c}{\textbf{SSDM-MRI}} \\
        \cmidrule(lr){3-11}
        &  & $\dfrac{\text{PSNR}}{\text{SSIM}}$  & $\dfrac{\text{PSNR}}{\text{SSIM}}$  & $\dfrac{\text{PSNR}}{\text{SSIM}}$ & $\dfrac{\text{PSNR}}{\text{SSIM}}$ & $\dfrac{\text{PSNR}}{\text{SSIM}}$ & $P^1$ & $\dfrac{\text{PSNR}}{\text{SSIM}}$ & $P^1$ & $\dfrac{\text{PSNR}}{\text{SSIM}}$ \\
        \midrule
        \multirow{4}{*}{\makecell{\textbf{Single-coil} \\ \textbf{brain (300 images)}}}
        & Gaussian 1D & 25.29 $\pm$ 1.612$^{**}$  &  25.16 $\pm$ 1.439$^{**}$  &  32.87 $\pm$ 2.563$^{**}$  &  31.74 $\pm$ 2.802$^{**}$  & 32.82 $\pm$ 1.987  & 0.018* & 32.94 $\pm$ 2.144  & 0.015* & \textbf{33.97 $\pm$ 1.974}  \\
        & \hspace{2em}8$\times$ & 0.81 $\pm$ 0.048$^{**}$ &  0.82 $\pm$ 0.061$^{**}$ &  0.91 $\pm$ 0.018$^{**}$ &  0.90 $\pm$ 0.039$^{**}$ &   0.92 $\pm$ 0.025 & 0.020* &  0.91 $\pm$ 0.031 & 0.016* &  \textbf{0.94 $\pm$ 0.021}  \\
        & Poisson 2D & 24.80 $\pm$ 1.981$^{**}$ & 24.49 $\pm$ 2.461$^{**}$ & 29.15 $\pm$ 2.360$^{**}$  & 28.94 $\pm$ 1.973$^{**}$  & 29.23 $\pm$ 2.340  & 0.016* & 29.11 $\pm$ 2.414  & 0.013* &\textbf{30.42 $\pm$ 2.144 }  \\
        & \hspace{2em}15$\times$ &  0.71 $\pm$ 0.060$^{**}$ &  0.68 $\pm$ 0.052$^{**}$ &   0.86 $\pm$ 0.038$^{**}$  &  0.85 $\pm$ 0.065 $^{**}$ &  0.87 $\pm$ 0.037 & 0.018* & 0.87 $\pm$ 0.039 & 0.015* & \textbf{0.90 $\pm$ 0.031} \\
        \midrule

        \multirow{4}{*}{\makecell{\textbf{Single-coil} \\ \textbf{knee (300 images)}}}
        & Gaussian 1D & 25.90 $\pm$ 2.038$^{**}$  &  25.61 $\pm$ 2.069$^{**}$ &  27.88 $\pm$ 2.336$^{**}$  & 27.50 $\pm$ 2.259$^{**}$  &  29.81 $\pm$ 2.152  & 0.089 & 29.63 $\pm$ 2.378  & 0.062 & \textbf{29.90 $\pm$ 2.121}   \\
        & \hspace{2em}8$\times$ & 0.77 $\pm$ 0.046$^{**}$&  0.75 $\pm$ 0.043$^{**}$ & 0.78 $\pm$ 0.042$^{**}$ &  0.79 $\pm$ 0.057$^{**}$ &   0.81 $\pm$ 0.072 & 0.017* &  0.81 $\pm$ 0.083 & 0.014* &  \textbf{0.83 $\pm$ 0.067}  \\
        & Gaussian 2D & 21.78 $\pm$ 1.749$^{**}$& 21.23 $\pm$ 1.341$^{**}$   & 27.61 $\pm$ 2.311$^{**}$ & 25.92 $\pm$ 2.872$^{**}$ &  28.14 $\pm$ 2.885 & 0.017* & 27.94 $\pm$ 2.693  & 0.012* & \textbf{28.48 $\pm$ 2.343}  \\
        & \hspace{2em}12$\times$ & 0.55 $\pm$ 0.082$^{**}$ &  0.53 $\pm$ 0.077$^{**}$ & 0.76 $\pm$ 0.024$^{**}$ & 0.74 $\pm$ 0.086$^{**}$ &  0.77 $\pm$ 0.022 & 0.016* &  0.77 $\pm$ 0.061 & 0.015* &  \textbf{0.80 $\pm$ 0.031}  \\
        \midrule

        \multirow{4}{*}{\makecell{\textbf{Multi-coil} \\ \textbf{brain (320 images)}}}
        & Gaussian 1D & 21.49 $\pm$ 1.214$^{**}$  & 20.75 $\pm$ 1.152$^{**}$   & 28.11 $\pm$ 2.261$^{**}$ & 27.66 $\pm$ 2.370$^{**}$ & 29.55 $\pm$ 1.905 & 0.016* & 29.67 $\pm$ 2.314  & 0.013* & \textbf{31.41 $\pm$ 1.924}   \\
        & \hspace{2em}12$\times$ &  0.62 $\pm$ 0.083$^{**}$ & 0.64 $\pm$ 0.079$^{**}$ & 0.85 $\pm$ 0.025$^{**}$ &  0.82 $\pm$ 0.029$^{**}$&  0.89 $\pm$ 0.046 & 0.017* &  0.88 $\pm$ 0.049 & 0.012* &  \textbf{0.91 $\pm$ 0.034} \\
        & Random 1D & 20.37 $\pm$ 1.947$^{**}$ &  19.88 $\pm$ 2.572$^{**}$ & 26.19 $\pm$ 2.161$^{**}$  & 25.47 $\pm$ 2.987$^{**}$  &  26.72 $\pm$ 2.087  & 0.016* & 26.78 $\pm$ 2.531 & 0.012* & \textbf{28.04 $\pm$ 1.941}  \\
        & \hspace{2em}12$\times$ &  0.53 $\pm$ 0.075$^{**}$ &   0.52 $\pm$ 0.065$^{**}$&  0.77 $\pm$ 0.062$^{**}$&  0.76 $\pm$ 0.082$^{**}$ &   0.85 $\pm$ 0.054 & 0.018* &  0.85 $\pm$ 0.061 & 0.015* &  \textbf{0.89 $\pm$ 0.047}  \\
        \midrule

        \multirow{4}{*}{\makecell{\textbf{Multi-coil} \\ \textbf{knee (320 images)}}}
        & Poisson 2D & 23.72 $\pm$ 2.181$^{**}$  & 22.52 $\pm$ 2.547$^{**}$   & 28.27 $\pm$ 2.444$^{**}$  & 28.01 $\pm$ 2.560$^{**}$ & 29.30 $\pm$ 2.385  & 0.016* & 29.13 $\pm$ 2.247  & 0.014* & \textbf{30.17 $\pm$ 2.148}   \\
        & \hspace{2em}15$\times$ &  0.63 $\pm$ 0.071$^{**}$  &  0.62 $\pm$ 0.062$^{**}$ &  0.74 $\pm$ 0.088$^{**}$ & 0.73 $\pm$ 0.077$^{**}$ &  0.77 $\pm$ 0.072 & 0.017* &  0.76 $\pm$ 0.079 & 0.013* &  \textbf{0.80 $\pm$ 0.069} \\
        & Gaussian 1D & 20.28 $\pm$ 2.749$^{**}$  & 20.19 $\pm$ 2.673$^{**}$  & 26.68 $\pm$ 2.797$^{**}$   & 26.45 $\pm$ 2.984$^{**}$ & 27.92 $\pm$ 2.679  & 0.017* & 28.04 $\pm$ 2.776  & 0.012* & \textbf{29.38 $\pm$ 2.531}   \\
        & \hspace{2em}12$\times$ &  0.56 $\pm$ 0.079$^{**}$ &  0.55 $\pm$ 0.077$^{**}$ &  0.72 $\pm$ 0.075$^{**}$ & 0.70 $\pm$ 0.084$^{**}$ &  0.77 $\pm$ 0.052 & 0.017* & 0.76 $\pm$ 0.071 & 0.013* &\textbf{0.81 $\pm$ 0.049}  \\
        \bottomrule
\end{tabular}}

\begin{tablenotes}
\footnotesize{
\item $^1P$-values were calculated using t-tests for comparison between the proposed SSDM-MRI and the other methods in terms of PSNR and SSIM, respectively. 
\item * indicates $P<$ 0.05 and ** indicates $P<$ 0.01 (note that $P$-values $<$ 0.01 are not displayed to save more space for better visualization)
}
\end{tablenotes}
\label{table2}
\end{threeparttable}
\end{table*}

Peak Signal to Noise Ratio (PSNR) and Structural Similarity (SSIM) \cite{wang2004image} were reported for quantitative assessments of different methods on magnitude and R2* images, while PSNR, High Frequency Error Norm (HFEN) \cite{langkammer2018quantitative}, and XSIM \cite{milovic2025xsim} (a specific SSIM implementation with optimization of internal parameters for evaluating QSM images, with hyperparameters set as follows: $L = 1$, $K_1 = 0.01$, and $K_2 = 0.001$), were reported for the QSM acceleration task. In addition, voxel-wise linear regressions on QSM values of five deep grey matter regions, i.e., Substantia Nigra (SN), Red Nucleus (RN), Globus Pallidus (GP), Caudate (CN), and Putamen (PU), were carried out to compare different methods. 
 
\section{Results}
\subsection{Evaluation on single-coil retrospective data}
Fig. \ref{fig3} compares SSDM-MRI with various MRI accelerating methods on two in-distribution fastMRI single-coil brain slices (AF = 12 and 8, respectively). Overall, SSDM-MRI showed the highest PSNR and SSIM on both datasets, and the minimum errors in the zoomed-in region. P-LORAKS and CS\textsubscript{PC} showed substantial residual artifacts (yellow arrows). All the other deep learning-based methods demonstrated over-smoothing and loss of fine details (red arrows) in the zoomed-in region, while the proposed SSDM-MRI better preserved the fine structures (green arrows).

SSDM-MRI's performance is evaluated on two OOD images (i.e., a fastMRI knee image and a pathological brain slice) in Fig. \ref{fig4}. SSDM-MRI resulted in the minimum error maps and the best fine image details for the knee slice, as pointed out by the green arrow, while all other methods showed apparent reconstruction errors. For the pathological brain image, SSDM-MRI led to the best metrics (30.31 PSNR and 0.95 SSIM) among all methods with the minimum error map in the zoomed-in region, while P-LORAKS and CS\textsubscript{PC} showed substantial artifacts (yellow arrows).

\subsection{Evaluation on multi-coil fastMRI data}
\subsubsection{Evaluation on retrospective data}

\
\newline
\indent The proposed SSDM-MRI is compared with P-LORAKS, CS\textsubscript{PC}, MoDL, U-net, and Score-MRI on two retrospective fastMRI brain (12 channels) and knee images (15 channels) in Fig. \ref{fig5} at AF = 12 and 15, respectively. P-LORAKS and CS\textsubscript{PC} led to apparent artifacts and reconstruction errors at these high AFs, while MoDL, U-net, and CCDF resulted in over-smoothed results. SSDM-MRI not only led to the best numerical metrics but also showed the best fine structures (pointed out by the green arrows).

Table \ref{table2} summarizes the performances of different methods on four retrospective dataset obtained from fastMRI: (1) single-coil brain, (2) single-coil knee, (3) multi-coil brain, and (4) multi-coil knee images at different AFs and different types of sampling masks. SSDM-MRI achieved on average the best PSNR and SSIMs for all four datasets, compared to the other algorithms.

\begin{figure*}[t]
\centering
\includegraphics[width=0.95\linewidth, scale = 0.8]{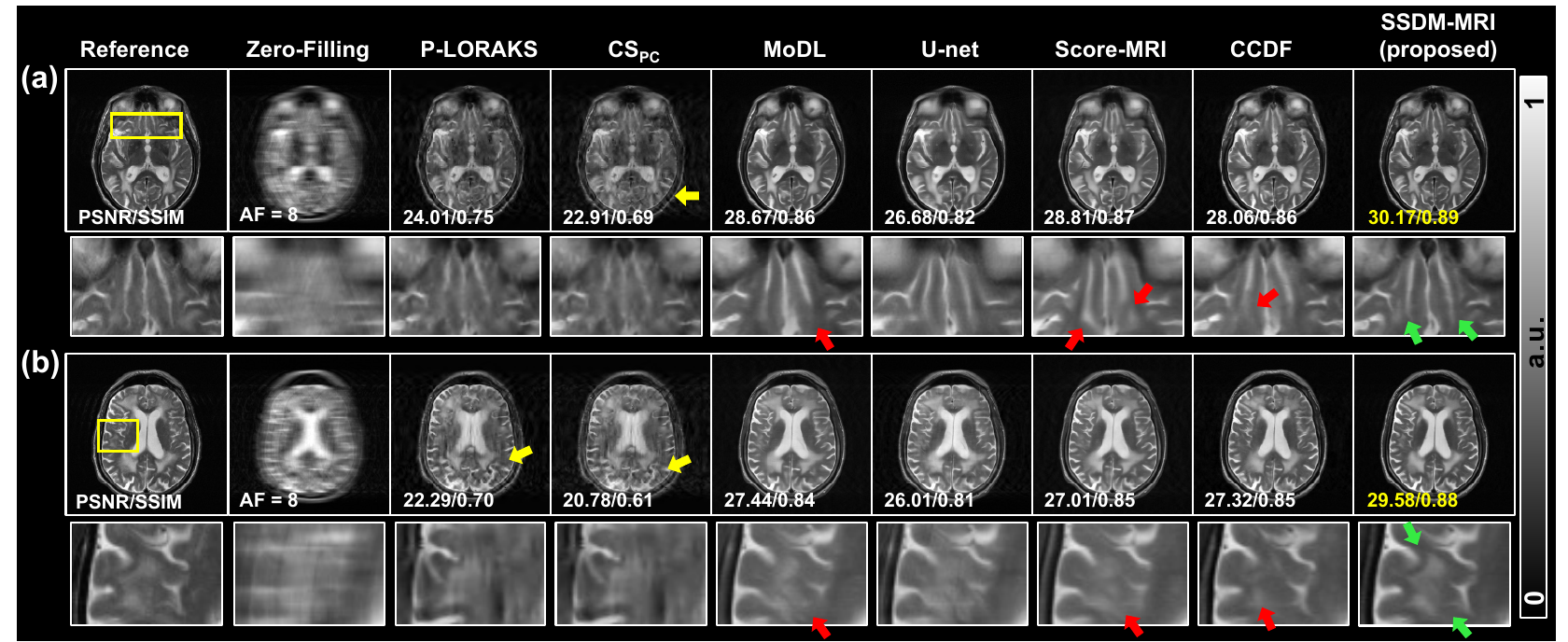}
\caption{Comparison of different reconstruction methods on two fastMRI multi-coil brain (16 receiver coils) data, prospectively undersampled at AF = 8. The yellow arrows point to the residual artifacts in P-LORAKS, CS\textsubscript{PC}, and the red arrows point to the over-smoothing and loss of fine structures in MoDL, Score-MRI, and CCDF. The green arrow points to better image fine details in SSDM-MRI. }
\label{fig6}
\end{figure*}



\begin{figure*}[htbp]  
\centering
\includegraphics[width=1\linewidth, scale = 0.95]{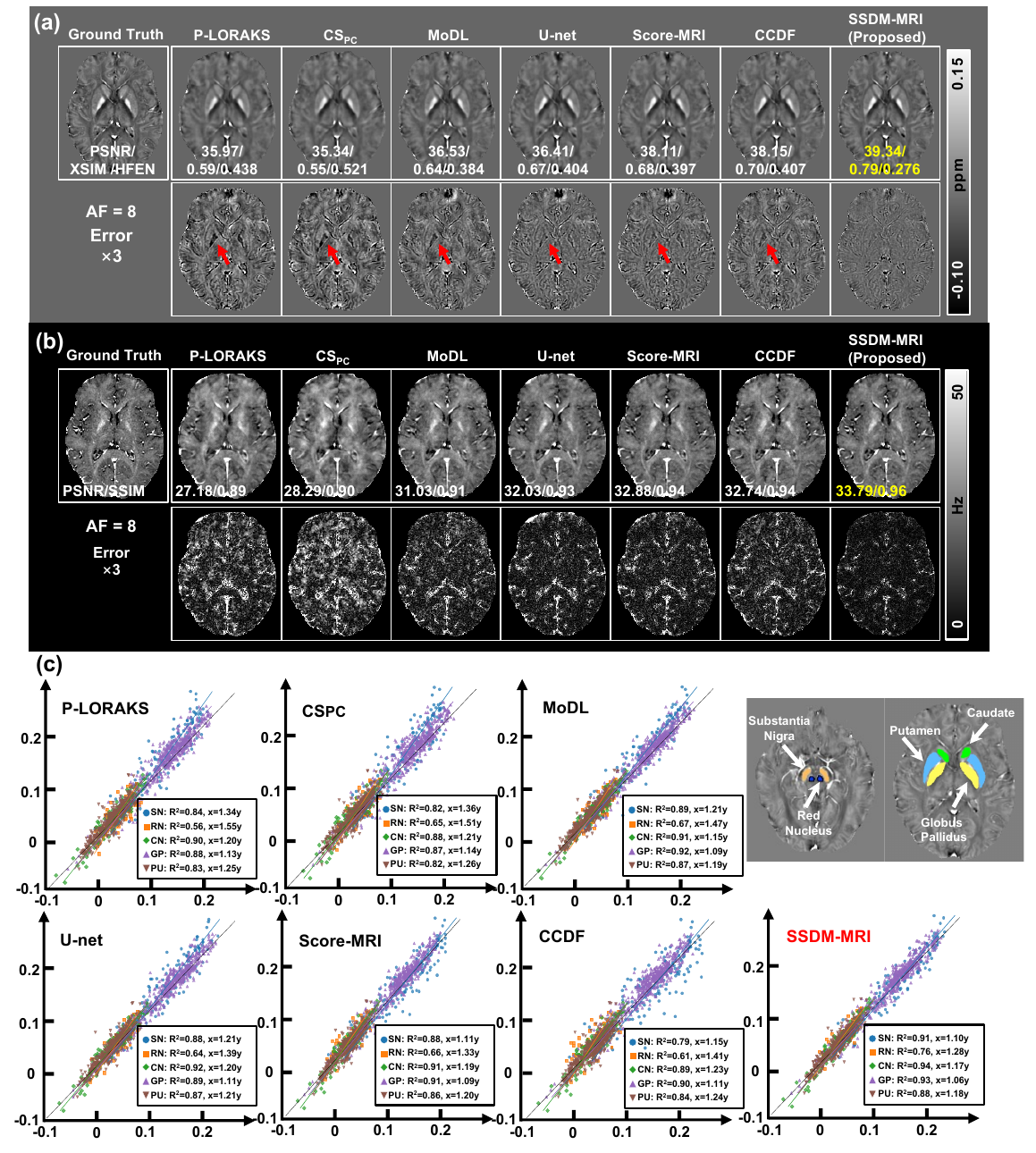}
\caption{Comparison of the proposed SSDM-MRI with other accelerating methods on one mGRE subject, retrospectively subsampled at AF = 8. (a) demonstrates the computed QSM images from the phase reconstructions and corresponding error maps. Red arrows point to apparent reconstruction errors in all methods except for the proposed SSDM-MRI. (b) shows the corresponding R2* results calculated from the mGRE magnitude images. (c) shows the scatter plots of susceptibility measurements in five deep grey matter regions, i.e., Substantia Nigra (SN), Red Nucleus (RN), Globus Pallidus (GP), Caudate (CN), and Putamen (PU), of various methods (x-axis) against the ground truth (y-axis). Linear regression results are reported correspondingly. }
\label{fig7} 
\end{figure*}

\subsubsection{Evaluation on prospective data}
\
\newline
\indent Different methods for two prospectively undersampled data are compared in Fig. \ref{fig6}. Similar to the simulation case, traditional algorithms, i.e., P-LORAKS, CS\textsubscript{PC} demonstrated susbtantial aliasing artifacts (yellow arrows), which were absent in the other methods. SSDM-MRI significantly reduced the image blurring apparent in U-net, MoDL, Score-MRI and CCDF in the zoomed-in region and demonstrated the best image fine details (green arrows). 

\subsection{Comparison on QSM acceleration task}
Fig. \ref{fig7} compares different methods on an mGRE brain volume at 8$\times$ acceleration. The proposed SSDM-MRI resulted in the best phase-based COSMOS-QSM results and the magnitude-based R2* images with the minimum error maps and the best quantitative metrics. Specifically, SSDM-MRI achieved the best PSNR, XSIM, and HFEN (39.34, 0.79, and 0.276) among all QSM reconstructions. Linear regression was also performed on the susceptibility values of various reconstruction methods against those of the ground truth, as shown in Fig. \ref{fig7}(c), which confirmed that the proposed SSDM-MRI achieved the most accurate quantitative measurements in the five deep grey matter regions ($R^2$ = 0.91, 0.76, 0.94, 0.93, and 0.88 for Substantia Nigra, Red Nucleus, Globus Pallidus, Caudate, and Putamen, respectively).

\begin{figure}[t]  
\centering
\includegraphics[width=0.95\linewidth,scale=1]{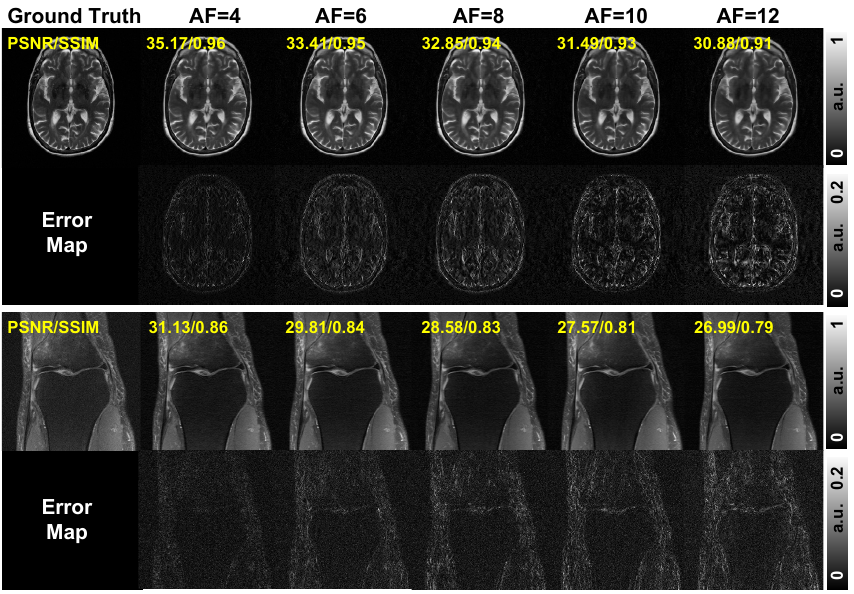}
\caption{Performance evaluation of the proposed SSDM-MRI against various accelerating rates from AF = 4 to 12, with PSNR and SSIM reported at the top part of the reconstructed brain and knee images. }
\label{fig8}
\end{figure}
\normalcolor

\subsection{The effects of accelerating factors}
The reconstructions of SSDM-MRI under 5 different AFs are shown in Fig. \ref{fig8}, on a retrospectively subsampled brain and a knee image. None of the reconstructed images showed apparent artifacts; however, as the AFs increases, more fine details gradually became blurred, which was also confirmed in the error maps. 

\vspace{-0.5em} 
\subsection{Reconstruction stability of SSDM-MRI}
\begin{figure}[b]  
\centering
\includegraphics[width=0.95\linewidth,scale=1]{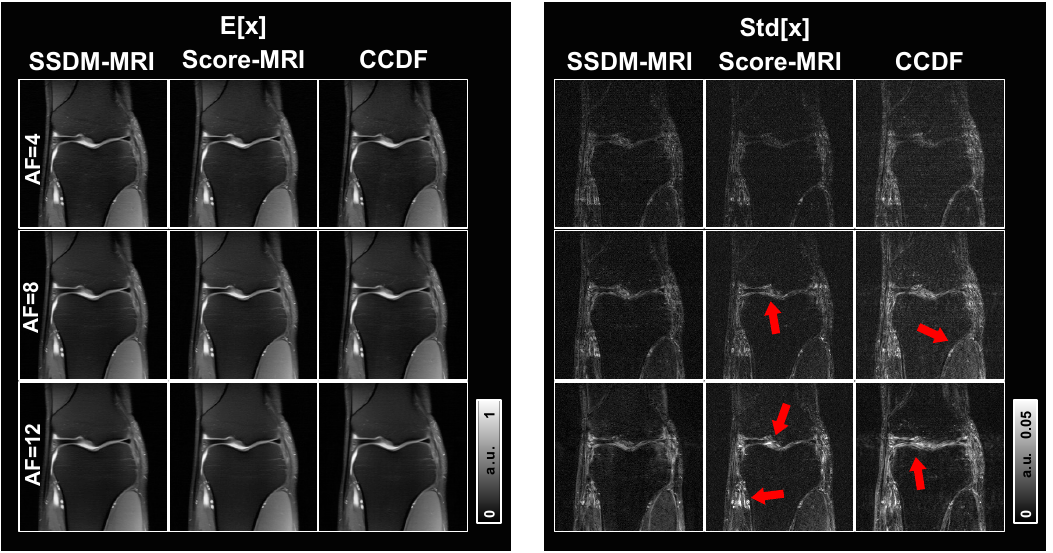}
\caption{The uncertainty maps of SSDM-MRI compared with Score-MRI and CCDF on one knee slice at accelerating rates of 4$\times$, 8$\times$, and 12$\times$, respectively. The left panel demonstrates the average results of 5 times repetitive reconstructions, while the right panel shows the corresponding standard deviations induced by the Gaussian noise in the inputs. Red arrows point to higher variations of multiple generations in Score-MRI and CCDF than the proposed SSDM-MRI. }
\label{fig9}
\end{figure}

Fig. \ref{fig9} illustrates the uncertainty maps of SSDM-MRI and Score-MRI of 5 repetitive reconstructions on one knee slice at AF = 4$\times$, 8$\times$, 12$\times$, respectively. As the AF increases from 4 to 12, all DM-based methods exhibited a notable increase in uncertainty images, with the Score-MRI and CCDF demonstrating slightly larger standard deviations than the proposed SSDM-MRI, as highlighted by the red arrows. Furthermore, a quantitative uncertainty analysis was also reported in Table \ref{table3} by calculating the mean and standard deviations of SSDM-MRI, Score-MRI, and CCDF over 10 repetitive reconstructions on a fastMRI brain image at AF = 8. SSDM-MRI achieved higher mean PSNR/SSIM values with smaller standard deviations than Score-MRI and CCDF, suggesting more stable and robust reconstructions. For example, the PSNR standard deviation of SSDM-MRI is only 0.355, compared to 0.512 and 0.634 for Score-MRI and CCDF, respectively.

\normalcolor
\subsection{Ablation study on data consistency, the proposed selective distillation algorithm, and the number of distillations}

\begin{figure}[t]  
\centering
\includegraphics[width=0.95\linewidth,scale=1]{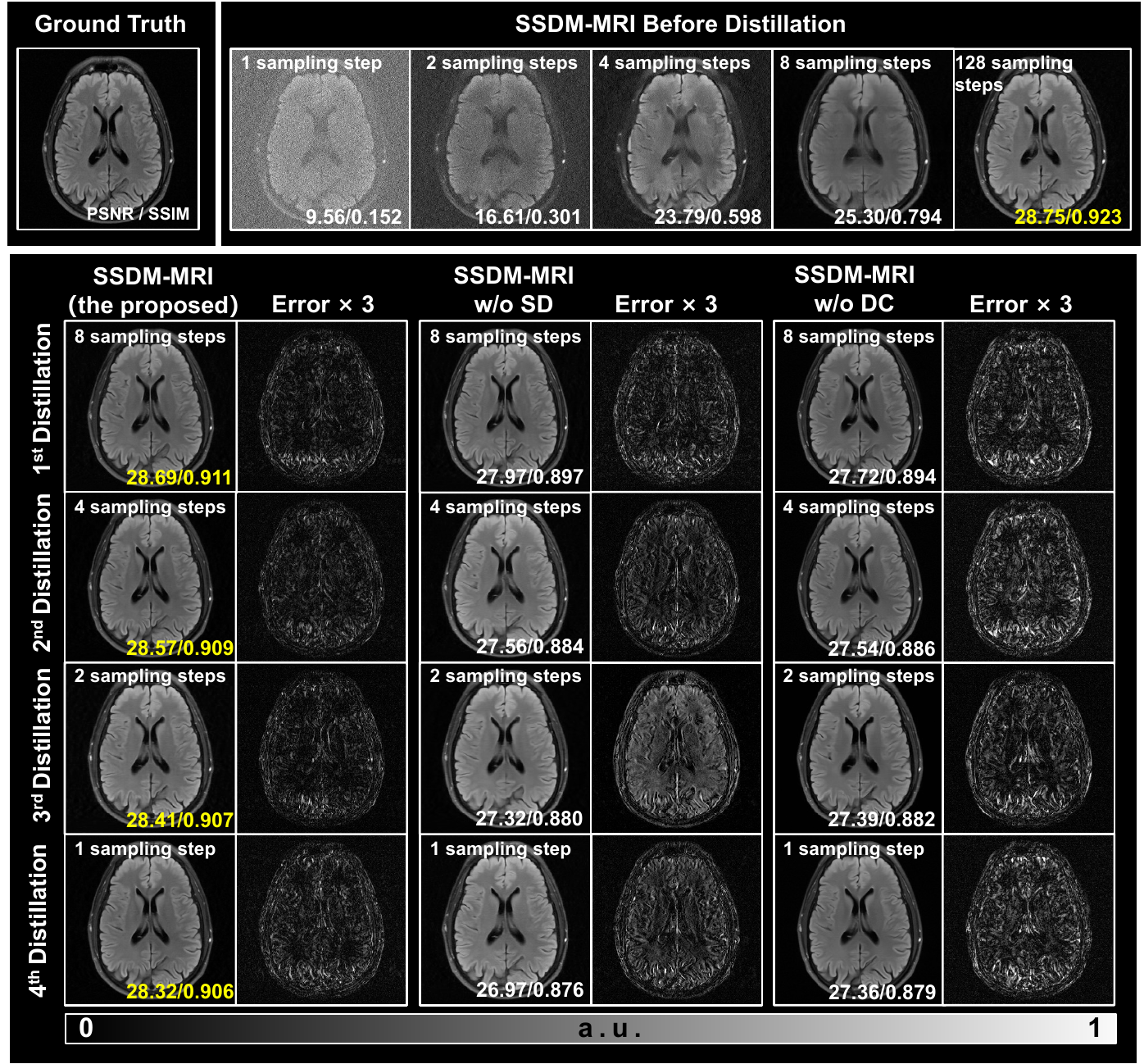}
\caption{Ablation illustration of the proposed SSDM-MRI results without data consistency (DC) or the selective distillation (SD) strategy (paired with the shortcut reverse sampling) at different distillation stages on a brain image, undersampled at AF = 10. SSDM-MRI before distillation is the pretrained conditional diffusion model; SSDM-MRI w/o SD is the SSDM-MRI trained using the conventional progressive distillation method without the proposed selective distillation design. }
\label{fig10}
\end{figure}
\normalcolor

\begin{table}[htbp]
\centering
\caption{Quantitative uncertainty analysis of the proposed SSDM-MRI compared with Score-MRI and CCDF by calculating the mean values and standard deviations of PSNRs and SSIMs of 10 repetitive reconstructions on a brain image undersampled at 8 $\times$. }

\begin{tabular}{ccc}
\toprule
\textbf{Method} & \textbf{PSNR} & \textbf{SSIM} \\
\midrule
\textbf{Score-MRI} & $32.43 \pm 0.512$ & $0.91 \pm 0.011$ \\
\textbf{CCDF}   & $32.21 \pm 0.634$ & $0.90 \pm 0.010$ \\
\textbf{SSDM-MRI (proposed)} & $\mathbf{33.52 \pm 0.355}$ & $\mathbf{0.92 \pm 0.006}$ \\
\bottomrule
\end{tabular}
\label{table3}
\end{table}

An ablation study was conducted to investigate the influences of data consistency (DC), the proposed selective distillation (SD) algorithm, and the number of distillations on the SSDM-MRI’s performance, as shown in Fig. \ref{fig10}. Overall, the performance of the proposed SSDM-MRI slightly decreases with increasing number of distillations, from the original 28.75 PSNR and 0.923 SSIM (128 reverse steps) to 28.32 PSNR and 0.906 SSIM (only 1 reverse step) after 4 distillation steps, which are still much better than the corresponding 1-step result before distillation (9.52/0.152 of PSNR/SSIM). Both DC operations and the proposed SD algorithms are important for SSDM-MRI to achieve high-quality reconstructions after distillation. For instance, distilling the pretrained conditional DM 4 times using the proposed iterative selective distillation with the incorporation of DC only observes a decrease of 0.43 and 0.017 in PSNR and SSIM, respectively, compared to a 1.78/0.047 PSNR/SSIM decrease for SSDM-MRI without Selective Distillation (i.e., SSDM-MRI using conventional progressive distillation), and a 1.39/0.044 PSNR/SSIM decrease for SSDM-MRI without data consistency. This improvement in alleviating the performance decrease after distillation might be due to the reason that the proposed iterative selective distillation algorithm focuses only on the most important parts of the reverse sampling, rather than the entire reverse path from the pure noise to clean images.

\section{Discussion and Conclusion}
In this work, we demonstrated the feasibility of MRI reconstructions from highly undersampled k-space with only a single-step posterior sampling of a diffusion model (DM). Specifically, an SSDM-MRI framework was constructed by iteratively distilling a pretrained conditional DM using the proposed iterative selective distillation algorithm (paired with a paired shortcut reverse sampling strategy). Extensive comparative studies were carried out to compare the proposed SSDM-MRI with multiple existing algorithms, i.e., P-LORAKS \cite{haldar2016p}, CS\textsubscript{PC} \cite{ong2018general}, MoDL \cite{aggarwal2018modl}, U-net \cite{ronneberger2015u}, and DM-based Score-MRI \cite{chung2022score} and CCDF \cite{chung2022come} on both public fastMRI images and a private multi-echo GRE (QSM) subject. In general, the proposed method exhibited the best results with the minimum reconstruction errors, the best numerical metrics and image fine details, and the most accurate susceptibility measurements in five deep grey matter regions (i.e., Substantia Nigra, Red Nucleus, Globus Pallidus, Caudate, and Putamen). 


There have already been several DM-based methods proposed for MRI acceleration \cite{chu2025highly,bian2024diffusion,chung2022score,cao2024high,liu2025score,chung2022diffusion,cui2024spirit,chen2025joint}, showing better reconstruction performances, especially when compared with traditional iterative and existing end-to-end deep learning (DL) based methods \cite{hyun2018deep,shan2023distortion,guo2023reconformer}. However, the computational costs of DM-based methods are usually much higher than previous DL methods, e.g., U-net-based frameworks, due to the need of a relatively large number of poster sampling steps, making the overall reconstruction time much longer and hindering their applications in clinical practice. To the best of our knowledge, there have not been any DM-based methods that can achieve high-quality MRI reconstructions in just one posterior sampling step, and SSDM-MRI should be the first one in the MRI community, and its reconstruction time is only comparable to end-to-end deep neural networks, as is evident in Table \ref{table1}. 

The proposed SSDM-MRI demonstrated promising generalization capability on the OOD knee data, though it was trained on brain images only, demonstrating its ability to learn the conditional prior underlaid in shared anatomical and statistical patterns of brain and knee images instead of simply memorizing the features. 

It is found in the ablation experiment (Fig. \ref{fig10}) that the performance of the proposed SSDM-MRI decreases with increasing number of distillations, which means that there is a compromise between the model’s performance and the computational cost. Reducing the necessary number of distillations might be an effective way to handle the trade-off between SSDM-MRI’s performance and the computational costs. In this work, an iterative selective distillation combined with a shortcut reverse sampling framework was proposed to reduce the number of distillations and, in the meantime, to focus only on the most important parts of the reverse sampling, rather than the whole reverse path from the pure noise to clean images, which is the key design for the improved performance. 

\begin{figure}[t]  
\centering
\includegraphics[width=0.95\linewidth,scale=0.9]{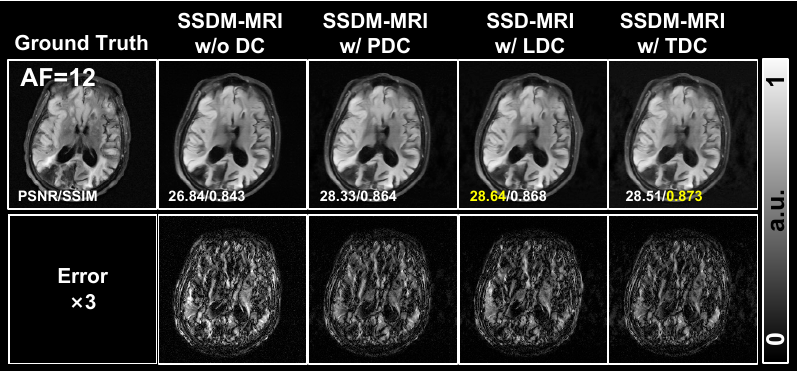}
\caption{Comparison of the proposed SSDM-MRI with different data consistency methods on a simulated brain image at an acceleration factor of 12. PDC is the conventional projection-based DC; LDC represents a learning-based DC method, and TDC denotes a two-step DC method. }
\label{fig11}
\end{figure}

\setlength{\tabcolsep}{2pt} 
\begin{table}[b]
\centering
\caption{The influences of three different DC methods on SSDM-MRI’s performances based on a fastMRI brain image dataset (128 single-coil images) retrospectively undersampled at 10 times. }
\begin{tabular}{lcccc}
\hline
 & \makecell{\textbf{SSDM-MRI}\\\textbf{w/o DC}} 
 & \makecell{\textbf{SSDM-MRI}\\\textbf{w/ PDC}} 
 & \makecell{\textbf{SSDM-MRI}\\\textbf{w/ LDC}} 
 & \makecell{\textbf{SSDM-MRI}\\\textbf{w/ TDC}} \\
\hline
\textbf{PSNR} & $29.60 \pm 1.801$ & $30.61 \pm 1.738$ & $\mathbf{30.78 \pm 1.701}$ & $30.68 \pm 1.634$ \\
\textbf{SSIM} & $0.859 \pm 0.040$ & $0.892 \pm 0.031$ & $0.897 \pm 0.028$ & $\mathbf{0.911 \pm 0.021}$ \\
\hline
\end{tabular}
\label{table4}
\end{table}

A relatively simple and straightforward projection-based DC (PDC) method was adopted in the current study after the reconstruction of the model to further improve the results and reduce the variance in the DM results. More advanced DC methods can be incorporated to further improve the results. We have conducted an experiment to compare the PDC method with two more advanced DC methods, i.e., a two-step DC (TDC) \cite{zheng2019cascaded}, and a learning-based DC (LDC) \cite{cheng2021learning} on retrospectively undersampled brain images, with the results reported in Fig. \ref{fig11} (one brain slice undersampled at 12$\times$) and Table \ref{table4} (128 brain images undersampled at AF = 8). Overall, SSDM-MRI with TDC or LDC achieved visually similar results and slightly better quantitative metrics (PSNR and SSIM), compared with SSDM-MRI using PDC. However, these improvements are much smaller than the improvements from SSDM-MRI without DC to SSDM-MRI with PDC. For example, as shown in Table \ref{table4}, changing PDC to LDC or TDC only results in an average 0.17/0.005 or 0.07/0.019 increase in PSRN/SSIM, respectively, compared to 1.01/0.033 increase from SSDM-MRI without data consistency to SSDM-MRI with the simplest PDC. In addition, PDC is more generic than LDC and TDC in some specific scenarios. For example, the LDC will need to be retrained to better suit different methods, while TDC cannot be directly used on the mGRE data reconstructions because it discards the very important phase component after the first DC step. Overall, the choice of DC methods can be very flexible if it is applied fairly to all comparative methods. 

\begin{figure}[t]  
\centering
\includegraphics[width=0.95\linewidth,scale=0.9]{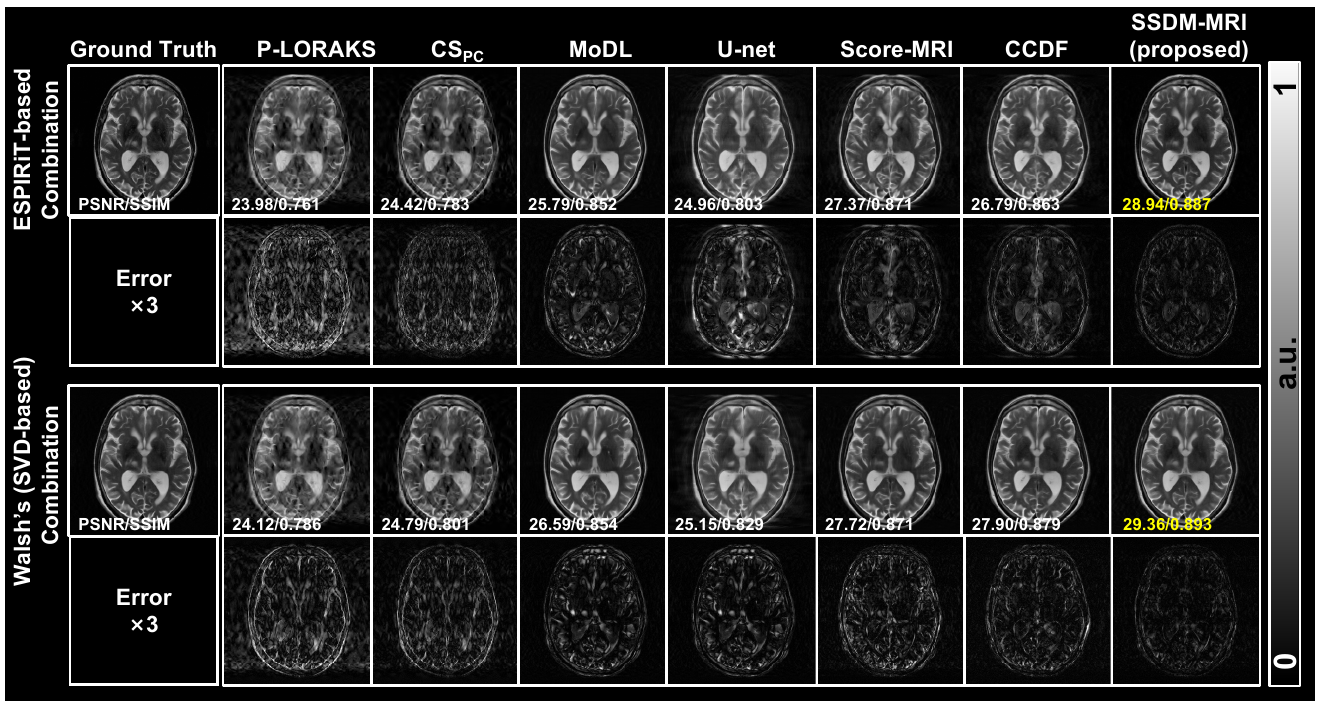}
\caption{Comparison of ESPIRiT and Walsh’s (SVD-based) method-based multi-coil reconstruction results of different MRI reconstruction methods on one brain image retrospectively subsampled at AF=10 (Gaussian 1D, ESPIRiT ACS = 6, image size = 320 $\times$ 320). }
\label{fig12}
\end{figure}

We also performed another experiment to compare multi-coil image reconstructions based on SVD and ESPIRiT coil combination in Fig. \ref{fig12}. It is obvious that the proposed SSDM-MRI achieved the best results compared to all other MRI reconstruction methods, regardless of which coil sensitivity estimation method is applied. In addition, it is demonstrated that the SVD-based method yields slightly higher PSNRs and SSIMs compared to the results obtained with the ESPIRiT method for all comparative methods. Therefore, the main conclusions will remain unchanged using either method. 

In the mGRE data-based experiment (i.e., the QSM acceleration experiment), different methods were repetitively tested on 1024 (128 slice number $\times$ 8 echo number) consecutive 2D images, covering the whole brain, before QSM and R2* reconstructions from the reconstructed phases and magnitude images, respectively. The proposed SSDM-MRI not only achieved on average the best results for all brain slices, but also preserved the best slice-to-slice coherence. It resulted in the best QSM measurements, demonstrating its capability of preserving latent and weak susceptibility signals buried in the MRI phase images. In addition, SSDM-MRI also achieved the most accurate results in R2*-based comparison, which further demonstrates its applicability in multi-echo imaging contexts. 

SSDM-MRI's performances were also evaluated against varying accelerating rates, and the results showed that no substantial artifacts were introduced. However, the image quality also gradually degraded with increasing AFs, and in the future, we will continue working on  obtaining high-quality images with high AFs. In addition, the reconstruction stability also needs to be further optimized for more reliable solutions. 

In conclusion, this study proposed a single-step diffusion model (DM)-based method, SSDM-MRI, for MR image reconstruction from highly undersampled k-space data, via a one-step posterior sampling from a conditional DM. Experimental results demonstrated that SSDM-MRI consistently led to improved results compared with several state-of-the-art methods on both magnitude and QSM reconstruction tasks, respectively, in a much shorter reconstruction time. 

\bibliographystyle{IEEEtran}
\bibliography{refs}
\end{document}